\documentclass[final,3p,times,twocolumn]{elsarticle}
\usepackage{graphicx}
\usepackage{nicefrac}
\usepackage{amsmath}
\usepackage{amssymb}
\usepackage{textcomp}
\usepackage{color}

\usepackage{epstopdf}
\biboptions{comma,sort&compress,numbers}
\journal{Carbon}
\begin{document}

\begin{frontmatter}

\title{Trace element content and magnetic properties of commercial HOPG samples studied by ion beam microscopy and SQUID magnetometry}

\author[exp]{D. Spemann}
\ead{daniel.spemann@gmx.de}

\author[exp]{P. Esquinazi\corref{corr1}}
\cortext[corr1]{Corresponding author. Tel/Fax: +49 341 9732751/69.
E-mail address: esquin@physik.uni-leipzig.de (P. Esquinazi)}

\author[exp]{A. Setzer}

\author[exp]{W. B\"{o}hlmann}

\address[exp]{Institute for Experimental Physics II, Universit\"{a}t Leipzig, Linn\'estr. 5, 04103 Leipzig, Germany}
\begin{abstract}

In this study, the impurity concentration and magnetic response of
nine  highly oriented pyrolytic
graphite (HOPG) samples with  different grades and from
different providers were determined using ion beam microscopy and
SQUID magnetometry. Apart from sideface contaminations in the
as-received state,
 bulk contamination of the samples in most cases consists of
disk-shaped micron-sized particles made of Ti and V with an
additional Fe contamination around the grain perimeter. The saturation
magnetization typically increases with Fe concentration, however,
there is no simple correlation between Fe content and magnetic
moment. The saturation magnetization of one, respectively six, out
of nine samples clearly exceeds the maximum contribution from pure
Fe or  Fe$_3$C. For most samples the temperature
dependence of the remanence  decreases linearly with $T$ -- a
dependence found previously for defect-induced magnetism (DIM) in
HOPG. We  conclude that apart from magnetic impurities,
additional contribution to the ferromagnetic magnetization exists
in pristine HOPG in agreement with previous studies. A comparative
study between the results of ion beam microscopy and the commonly
used EDX analysis shows clearly that EDX is not a reliable method
for quantitative trace elemental analysis in graphite, clarifying
weaknesses and discrepancies in the element concentrations given
in the recent literature.
\end{abstract}
\begin{keyword}
HOPG\sep trace element content\sep magnetic properties
\end{keyword}
\end{frontmatter}
\section{Introduction}\label{Intro}
In the last years the possibility to have magnetic order in different kinds of solids above room temperature without nominally magnetic ions, like the usual transition or rare earth elements, has attracted the attention of the solid state community. Although some theoretical and experimental works in the past provided some hints for the existence of this apparently unusual phenomenon, it has been only recently that we became aware that different kinds of defects, like vacancies, hydrogen or a combination of those with nominally non-magnetic elements can trigger magnetic order in solids \cite{sto10,and10,vol10,yaz10,dim13}.

This phenomenon, named defect-induced magnetism (DIM), has been mostly studied, theoretically and experimentally, in the single element graphite/graphene. More than ten years ago, systematic studies of the magnetic properties of different graphite samples with different magnetic impurity concentrations suggested that an extra magnetic contribution, other than from impurities, should exist \cite{yakovjltp00,pabloprb02}. In general, the small ferromagnetic moment observed in commercial, as-received highly oriented pyrolytic graphite (HOPG) samples makes the detailed knowledge of the contribution from magnetic impurities imperative to understand its origin. In this work, we have studied HOPG samples with different grades from three different commercial sources, i.e. a total of nine HOPG samples. This study, therefore, gives a fairly reproducible spectrum on the different magnetic contributions in the HOPG samples available nowadays.

An accurate measurement of impurity concentrations in the ppm range and in micrometer small grains is not simple and only possible with an experimental method which provides elemental imaging with excellent detection limits in the ppm and sub-ppm range together with reliable quantification, preferably in a non-destructive way. The method used in this work, Particle Induced X-ray Emission (PIXE), has the necessary requirements for this kind of studies. In this work, we show that a large contamination with magnetic elements is found at the sidefaces of as-received HOPG samples, very probably originating from the cutting of the sample prior shipping. The observed amount of the sideface impurities can overwhelm by far the usually observed impurity concentration in the bulk and, if no thorough sample cleaning is done, clearly prevents the measurement of the magnetic contribution from DIM in pristine HOPG samples.

On the other hand, the sole measurement of the magnetic moment of a sample with a known amount of impurities does not provide always with a clear statement, whether the ferromagnetism is or is not due to impurities \cite{Esquinazi1156,Sepioni47001,Spemann57006,Venkatesan279}. In general and  for small impurity concentration and grains \cite{Esquinazi1156}, only an upper estimate of the ferromagnetic signal from those grains can be simply done. This is due to the fact that not only the grain size, but also the, in general difficult to quantify, stoichiometry and structure of the existing ferromagnetic phases determine the magnetic signal. Therefore, demonstrated for the case of graphite, we will show in this work that together with a careful impurity measurement, the temperature dependence of the remanent magnetic moment helps to discern, which magnetic contributions are active in the sample.

The manuscript is divided in three more sections and a conclusion. In the next section~\ref{exp} we provide details on sample preparation, the used elemental analysis methods and the commercial Superconducting Quantum Interferometer Device (SQUID) used for magnetometry. Taking into account the, in general, limited knowledge on the possibilities of PIXE in contrast to usual methods for elemental analysis like Scanning Electron Microscopy combined with Energy Dispersive X-ray analysis (SEM/EDX) \cite{Sepioni47001,Spemann57006,Venkatesan279}, the manuscript includes a complete trace element analysis obtained with PIXE in section \ref{te} and a comparison with the results obtained with EDX analysis in section~\ref{c-edx}. Section \ref{sf} discusses the impurity distribution at the sidefaces of the commercial HOPG, a fact that we believe is of importance for any future discussion on the ``intrinsic'' bulk impurity concentration presented in section~\ref{bc}. An elemental analysis of single metallic grains found in HOPG samples is given in section~\ref{PIXEgrains}. In section~\ref{mp} we present and discuss the magnetic characterization done with the SQUID in all HOPG samples and compare it with the information obtained from the elemental analysis. The conclusion is given at the end of the manuscript.

\section{Experimental}\label{exp}
\subsection{Sample preparation}
The samples studied are commercially available HOPG from Advanced Ceramics (now Momentive Performance Materials), NT-MDT and SPI Supplies \cite{hopg}. From each of these companies, samples were purchased in the three available structural grades designated as ZYA, ZYB and ZYH. In case of SPI Supplies the corresponding designation is SPI-1, -2 and -3. In the following, the samples are named by the company (Advanced Ceramics is abbreviated as AC) and the structural grade, e.g. AC ZYA denotes the ZYA-grade sample from Advanced Ceramics.

The HOPG samples were wire-cut into pieces of $(5\times 5)$~mm$^2$ size each and thoroughly cleaned in an ultrasonic bath with ethanol for SQUID magnetometry. On one piece of ZYA-grade HOPG from each company trace elemental analysis was performed in the as-received state, i.e. without sample cleaning to check for possible contaminations, especially at the sidefaces. For this purpose, the samples were only cleaved with a CuBe-knive and glued on Si substrates using varnish with the cleaved surface on top. In order to characterize the ``intrinsic'' bulk trace element content using PIXE and SEM/EDX analysis, the samples were additionally cleaned three times with ethanol in an ultrasonic bath, glued on Si substrates using varnish and the top surface removed by tape stripping to prepare a fresh one onto which the ion/electron beam was directed.

\subsection{PIXE and RBS ion beam microscopy}\label{PIXEexp}
Trace elemental analysis of the samples was performed at the LIPSION facility of the University of Leipzig \cite{Butz323} with PIXE \cite{PIXEbook} and Rutherford Backscattering Spectrometry (RBS) \cite{RBSbook} using a 2.28~MeV proton microbeam focused to $1-2$~$\mu$m diameter. The proton microbeam was raster-scanned across the sample surface and the characteristic X-rays and backscattered protons simultaneously recorded for each scan pixel. In this way, non-destructive quantitative imaging of the elemental content is possible with micron lateral resolution. Detailed information about the LIPSION facility can be found in \cite{Spemann2175}.

In contrast to the commonly used EDX analysis with its comparably poor minimum detection limits (MDL), PIXE allows true trace elemental analysis in carbon with MDLs $\lesssim 0.1$~$\mu$g/g for 3d-elements like Fe. A comparison with neutron activation analysis showed that Fe concentrations as low as 0.17~$\mu$g/g can be accurately determined using the PIXE method \cite{Esquinazi1156}. In addition, MeV protons penetrate much deeper into the material than electrons of several tens of keV. According to SRIM-2013 simulations \cite{Ziegler1818} 2.28~MeV protons penetrate 47~$\mu$m deep into graphite where they generate 90\% of the total X-ray yield from Fe atoms within the first 27~$\mu$m \cite{Ryan170}. At this depth the beam diameter has increased by only $\approx 1.4$~$\mu$m due to scattering processes, i.e. the proton microbeam is still well-focused allowing PIXE to be used as a bulk-sensitive analysis technique with good imaging capabilities even for trace elements buried several microns below the graphite surface. Since the X-ray emission process used in PIXE relies on the ionization of inner-atomic shells (in case of Fe the innermost K-shell is used) which are practically unaffected by the chemistry of the sample, PIXE can be considered as being ``chemically blind'' as are EDX or XRF. As a consequence, the actual distribution of impurity atoms does not affect the X-ray production from a certain amount of these impurities, no matter whether they are homogeneously distributed or enriched in small grains. Since ion channeling can be excluded in our measurements, the lattice site location of impurity atoms does not influence their detection efficiency by the PIXE method as well. This ensures that all impurity atoms can be detected, no matter in which chemical state they are or how they are distributed within the volume probed by the ion beam -- an important prerequisite for reliable quantitative elemental analysis.

The PIXE spectra were recorded using a high-purity GUL0110 Germanium detector from Canberra with an active area of 95~mm$^2$ subtending a solid angle of 150~msr and an energy resolution of 144~eV at 5.9~keV. The spectra were analyzed using GeoPIXE II \cite{Ryan170}. Whereas for the calculation of bulk concentrations, graphite with a thickness greater than the proton range was used as matrix in the data analysis, a thin layer of Fe with a mass thickness of 0.1~mg/cm$^2$ was assumed for the analysis of the sideface contamination. This accounts for the negligible energy loss of the protons and x-ray absorption in the thin surface layer of contamination at the sideface of the samples. The thickness itself is arbitrarily chosen and cancels out in the calculation of the mass/area value of the contamination.

\begin{figure}[ht!]
\centering
    \includegraphics[width=1\columnwidth]{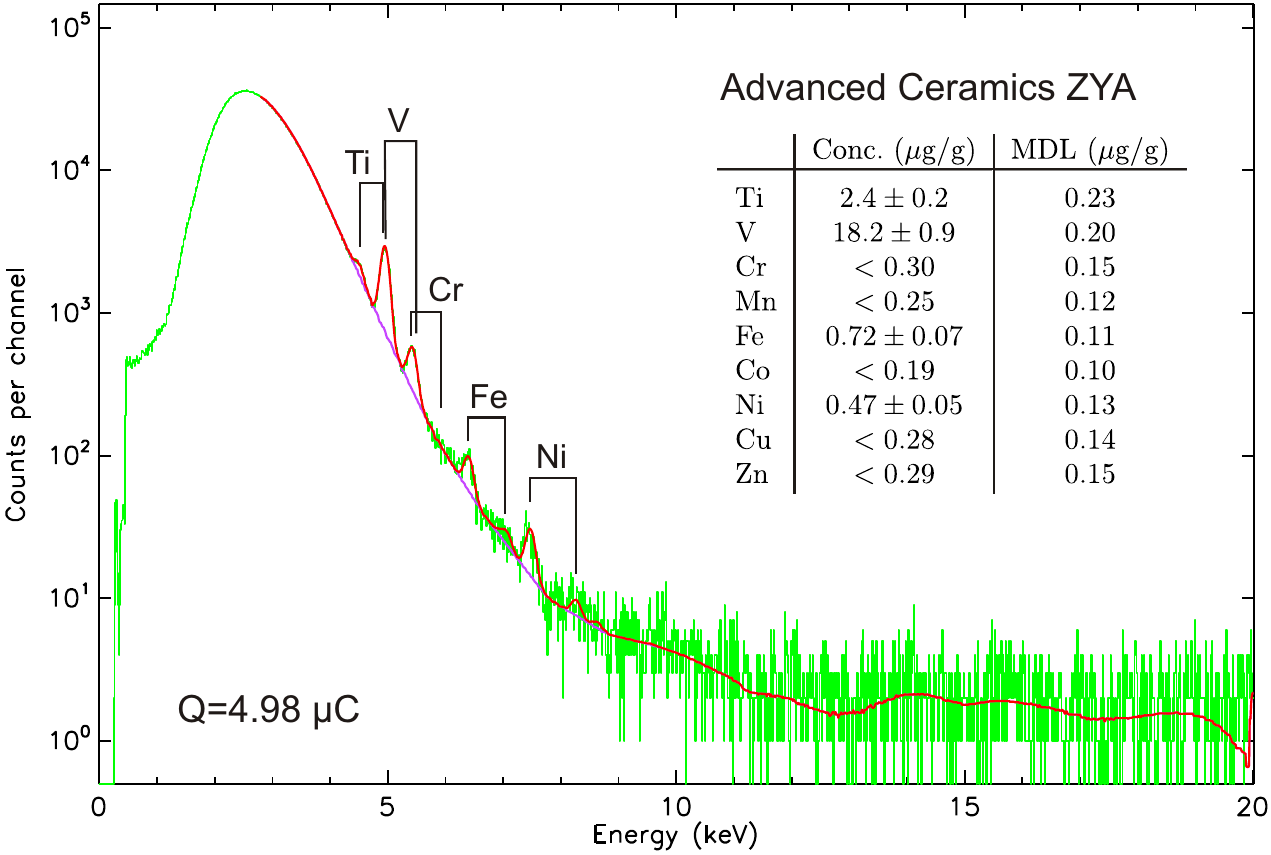}
    \caption{PIXE spectrum from ZYA-grade HOPG from Advanced Ceramics recorded with a collected proton charge
             of $Q=4.98$~$\mu$C. The green curve are the measured data, the violet and red curve are the background simulation and fit to the data, respectively, from GeoPIXE II. The extracted concentrations as well as minimum detection limits are given and the corresponding X-ray lines for the detected elements indicated.}
    \label{PIXEspec}
\end{figure}

Figure \ref{PIXEspec} shows a typical PIXE spectrum from ZYA-grade HOPG from Advanced Ceramics together with the extracted elemental concentrations.

Whereas PIXE provides excellent sensitivity, but no direct depth profiling capabilities, RBS inherently allows depth profiling of element concentrations if they are sufficiently high. As will be shown later, RBS can be used to determine the thickness of metallic particles and their location below the surface in a non-destructive way, i.e. without the need to prepare cross-sections from the sample for EDX or TEM analysis. Furthermore, RBS was used in this study to accurately determine the applied proton charge from the RBS yield of the carbon bulk.

The RBS spectra were recorded using an annular PIPS detector from Canberra with an area of 275~mm$^2$, an effective backscattering angle of $172^\circ$, a solid angle of 86~msr and an energy resolution of 10.6~keV for 2.28~MeV protons. Afterwards, the spectra were analyzed using XRUMP \cite{Doolittle344}.

\subsection{SEM and EDX analysis}
For comparison with previously published studies \cite{Sepioni47001,Venkatesan279} and ion beam microscopy selected samples were further analyzed with SEM and EDX using the Dual Beam Microscope Nova NanoLab 200 from FEI Company. Prior elemental analysis of single metallic particles, backscattered electron (BE) imaging was used to locate them due to the $Z$-contrast between the carbon bulk and the heavier 3d-elements of the grains. Then EDX spectra and element maps were recorded using 20~keV electrons and analyzed with the EDAX software.

\subsection{SQUID magnetometry}
Magnetization measurements were performed with a SQUID magnetometer MPMS-7 from Quantum Design with Reciprocal Sample Option (RSO) and the magnetic field applied parallel to the graphene planes of the samples (within an experimental resolution of $\pm 3^\circ$). Several years of experience in measuring graphite samples with SQUID  \cite{barzola2,Esquinazi1156,ram10} and the excellent reproducibility of the used apparatus allows us a sensitivity of  $\lesssim 2 \times 10^{-8}~$emu.

\section{Trace element content in HOPG}\label{te}
\subsection{Sideface contamination of as-received samples}\label{sf}
\begin{figure*}[ht!]
\centering
    \includegraphics[width=0.8\linewidth]{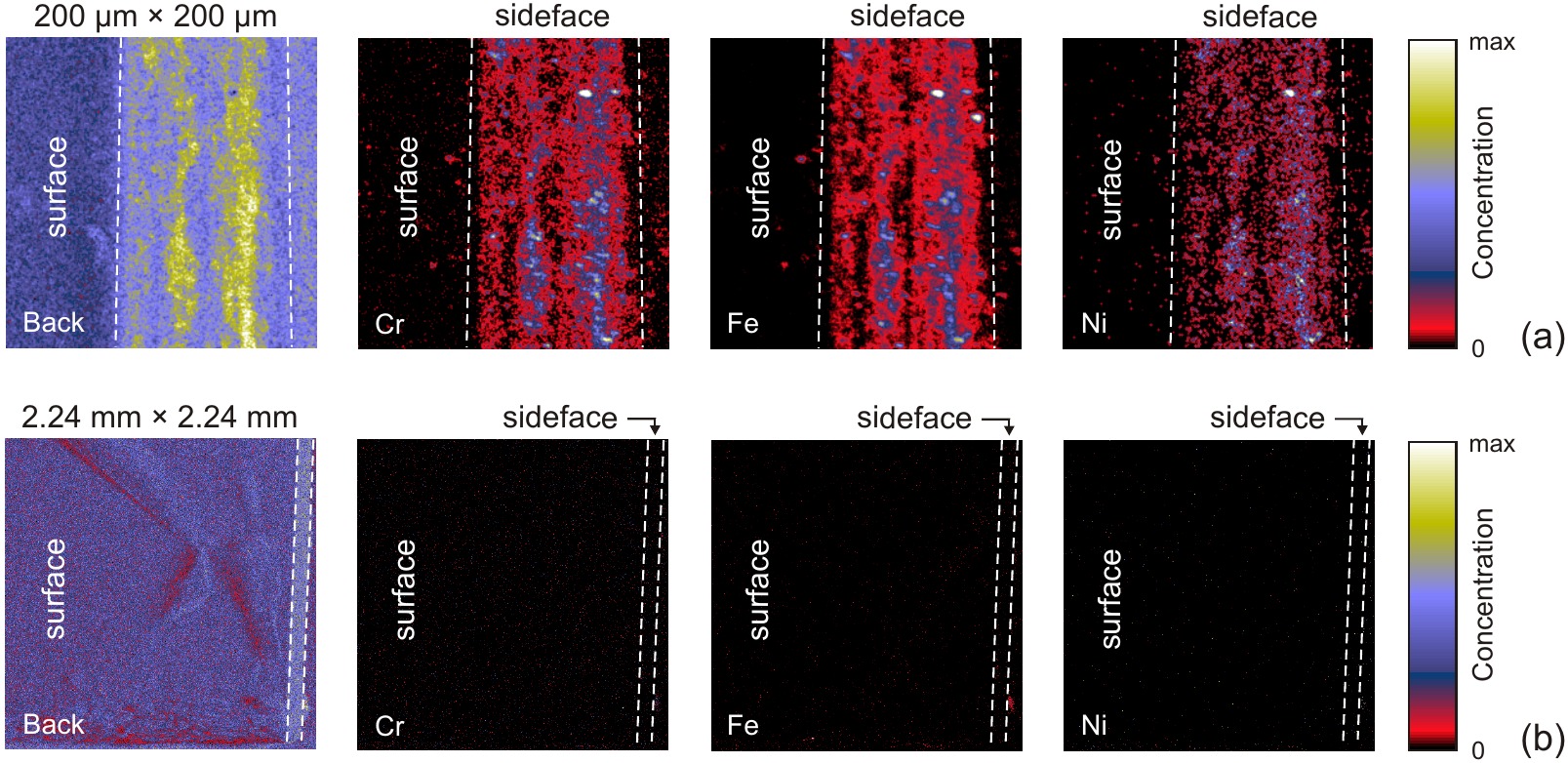}
    \caption{X-ray yield in false color scale for the background and the elements Cr, Fe and Ni: (a) SPI-1 sample in the as-received state ($200~{\rm \mu m} \times 200~{\rm \mu m}$ scan size, $Q=0.87$~$\mu$C applied proton charge) showing a severe sideface contamination with Cr, Fe and Ni, probably originating from a stainless steel tool used for sample cutting. Some loose particles of contamination can even be found on the top surface close to the sideface; (b) SPI-1 after ultrasonic cleaning ($2.24~{\rm mm} \times 2.24~{\rm mm}$ scan size, $Q=5.24$~$\mu$C applied proton charge) with the same maximum concentration values of the color scale as in (a). As can be seen between the two dashed lines, the sideface is not contaminated anymore.}
    \label{sideface}
\end{figure*}

In order to check for contaminations in the as-received state the HOPG samples of ZYA grade were analyzed with the proton beam incident on both top surface and sideface with a tilt angle of $15^\circ$ between beam and sample normal. The yield from the secondary electron background (see Fig.~\ref{PIXEspec} for comparison) was used to differentiate between both areas designated as ``surface'' and ``sideface'' in Fig.~\ref{sideface}
where the X-ray yields, i.e. elemental concentrations, are displayed in false color scale for the SPI-1 sample. As can be seen in Fig.~\ref{sideface}(a), the sideface is strongly contaminated with Cr, Fe, and Ni, all three showing an identical distribution indicating that they originate from the same source. The quantitative analysis of this contamination reveals a composition of 16.0\% Cr, 77.2\% Fe and 6.8\% Ni by weight which fits with the frequently used non-magnetic, austenitic SAE grade 301 stainless steel \cite{SAEsteel}. It is therefore reasonable to assume that this contamination originates from the cutting of the samples prior shipping using a stainless steel tool.

Contaminations of similar distribution and composition can be found on the sidefaces of AC ZYA and NT-MDT ZYA samples as well, however, to a less severe degree. As Tab.~\ref{Bulk_conc} shows, {702~ng Fe per cm$^2$ sideface area} was found for SPI-1, whereas for the AC and NT-MDT samples Fe concentrations amount to 15.8~ng/cm$^2$ and 155~ng/cm$^2$, respectively. In addition, Cl, K and Ca contaminations were detected on the sidefaces of AC ZYA and SPI-1 with distributions different from Cr, Fe and Ni. Even though they were not measured it is reasonable to assume that the sidefaces of the ZYB and ZYH samples are contaminated as well in the as-received state.

After thorough ultrasonic cleaning, however, the sidefaces are free of contaminants. In order to check this, the large scan from which the bulk trace element content was determined was placed such that the sideface area was included as well, see Fig.~\ref{sideface}(b). Clearly, the sideface does not show enhanced concentrations of Cr, Fe and Ni anymore compared to the top surface. We conclude that ultrasonic cleaning is mandatory prior use of these HOPG samples in contamination-critical applications.

\subsection{Bulk concentrations of trace elements}\label{bc}
Due to the excellent sensitivity of PIXE, trace element analysis can be performed by simply scanning the proton beam over a large area of the sample and collecting the X-ray photons from the trace element atoms. For this purpose, a $2.24~{\rm mm} \times 2.24~{\rm mm}$ sized scan was made covering a substantial portion of the whole sample area. The PIXE spectra (see as example Fig.~\ref{PIXEspec} for the AC ZYA sample) were analyzed and the concentrations extracted are given in Tab.~\ref{Bulk_conc}.
\begin{table*}
\begin{tabular}{|l|c|c|c|c|c|c|c|}
\hline
\rule{0mm}{3.5mm}Sample&Sideface&\multicolumn{6}{|c|}{Concentrations in HOPG bulk ($\mu$g/g)}\\
~&Fe (ng/cm$^2$)&Ti&V&Cr&Fe&Co&Ni\\ \hline
\rule{0mm}{3.5mm}AC ZYA&$15.8\pm 1.1$&$2.4\pm 0.2$&$18.2\pm 0.9$&$<0.30$&$0.72\pm 0.07$&$<0.19$&$0.47\pm 0.05$\\
AC ZYB&n.d.&$4.9\pm 0.4$&$10.5\pm 0.5$&$<0.32$&$<0.23^*$&$<0.21$&$<0.28$\\
AC ZYH&n.d.&$1.9\pm 0.2$&$24.5\pm 1.2$&$<0.30$&$22.6\pm 1.1$&$<0.25$&$4.2\pm 0.3$\\
NT-MDT ZYA&$155\pm 8$&$<0.50$&$<0.40$&$<0.28$&$0.55\pm 0.05$&$<0.20$&$<0.27$\\
NT-MDT ZYB&n.d.&$12.6\pm 0.9$&$0.80\pm 0.10$&$<0.29$&$10.4\pm 0.5$&$<0.23$&$<0.31$\\
NT-MDT ZYH&n.d.&$5.6\pm 0.3$&$1.40\pm 0.14$&$0.30\pm 0.04$&$10.2\pm 0.5$&$<0.22$&$<0.28$\\
SPI-1&$702\pm 35$&$<0.50$&$<0.40$&$<0.27$&$0.66\pm 0.06$&$<0.19$&$<0.27$\\
SPI-2&n.d.&$1.5\pm 0.2$&$<0.40$&$<0.31$&$9.4\pm 0.5$&$<0.24$&$<0.31$\\
SPI-3&n.d.&$2.1\pm 0.2$&$5.4\pm 0.5$&$<0.33$&$8.4\pm 0.4$&$<0.23$&$<0.33$\\ \hline
\multicolumn{8}{l}{\rule{0mm}{3mm}{\small n.d.: not determined; $^*$ Fe concentration in a similar AC ZYB sample: $(0.17\pm 0.03)~\mu$g/g \cite{Esquinazi1156}}}\\
\end{tabular}
\caption{Sideface contamination with Fe and bulk concentrations of trace elements in HOPG. With the exception of AC ZYB the ZYA samples with the highest structural quality have the lowest Fe concentrations $<1~\mu$g/g. In addition to the elements listed the concentrations of Mn, Cu and Zn were determined for all samples as well and found to be $<0.28~\mu$g/g for Mn, $<0.33~\mu$g/g for Cu and $<0.34~\mu$g/g for Zn.}
\label{Bulk_conc}
\end{table*}

The findings can be summarized as follows: (i) the ZYA-grade samples have Fe concentrations below 1~$\mu$g/g; (ii) the ZYB and ZYH samples of NT-MDT and SPI all have similar Fe concentrations of the order of 10~$\mu$g/g, whereas the AC ZYB has the lowest Fe contamination of $<0.23~\mu$g/g ($(0.17\pm 0.03)~\mu$g/g was determined for a similar sample a few years ago \cite{Esquinazi1156}) and the AC ZYH the highest Fe concentration of 22.6~$\mu$g/g; (iii) most of the samples also contain Ti and V with varying concentrations between the different samples.

From the concentrations it is immediately clear that the contamination is not due to stainless steel particles as was the case for the sidefaces. Nevertheless, it is interesting to compare the Fe contamination of the bulk and sideface. In case of the $5\times 5$~mm$^2$ SPI-1 sample the sideface contains twice as much Fe than the bulk, which again illustrates the importance of ultrasonic cleaning.

We would like to point out that, with the exception of AC ZYA and AC ZYB samples, we have not analyzed any of the other samples before and can therefore not make any statement about the variation of the concentration values between different batches of these samples. In \cite{Venkatesan279} the Fe concentration in a SPI-2 sample determined by Instrumental Neutron Activation Analysis (INAA) was almost three times higher than in our sample. In that publication, however, the authors did not provide any information on sample cleaning.

AC ZYA samples have been used in our studies on DIM and therefore frequently analyzed in the last ten years. We found that their trace element content shows little variation between different batches. In case of AC ZYB one sample was analyzed a few years ago \cite{Esquinazi1156}, again showing similar trace element concentrations as given in Tab.~\ref{Bulk_conc}.

\begin{figure}[ht]
\begin{center}
\includegraphics[width=1\columnwidth]{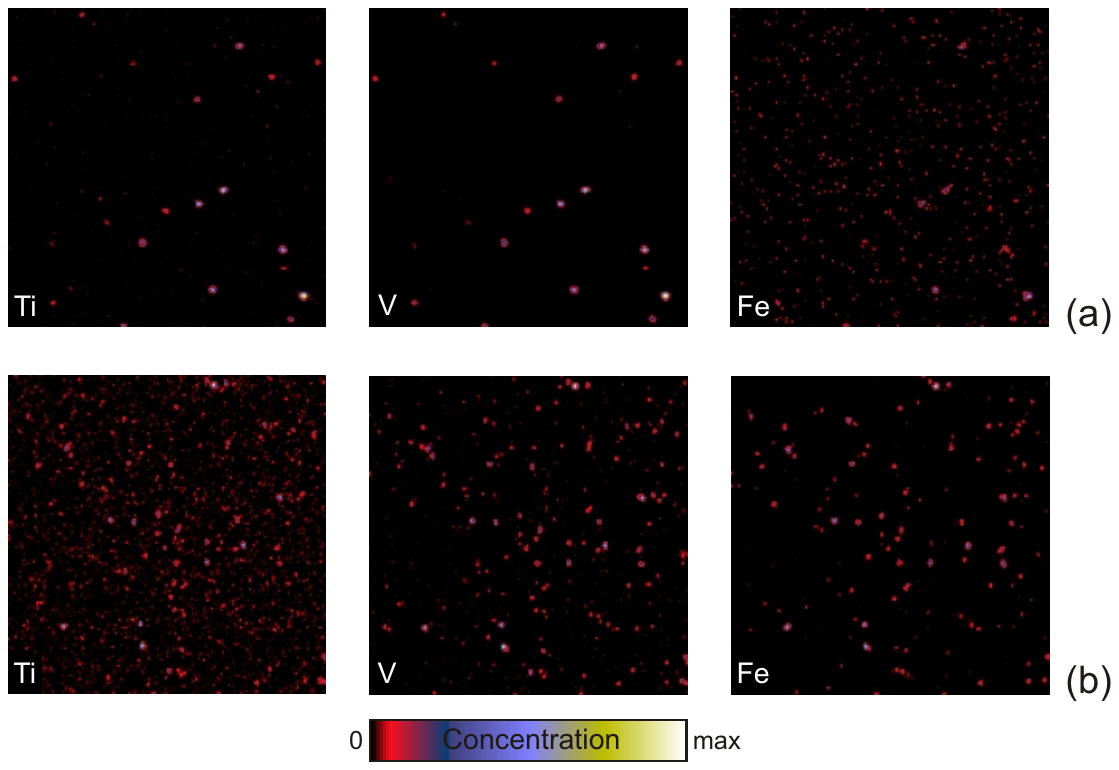}
\caption[]{Ti, V and Fe distribution in a $400~{\rm \mu m} \times 400~{\rm \mu m}$ sized scan area for (a) AC ZYA and (b) SPI-3. Ti and V are strongly correlated and show the same distribution. Most of the (Ti,V) grains contain Fe, but some do not. In addition, some Fe grains do not contain Ti or V.}
\label{PIXE_overview}
\end{center}
\end{figure}
As already stated in \cite{Esquinazi1156} and confirmed recently in \cite{Sepioni47001,Venkatesan279} the
trace elements are not distributed homogeneously within the sample, but located in micrometer large grains
that are homogeneously dispersed within the bulk. This is illustrated in Fig.~\ref{PIXE_overview} where the Ti, V and Fe maps of a $400~{\rm \mu m} \times 400~{\rm \mu m}$ scan are shown for the AC ZYA and SPI-3 sample in (a) and (b), respectively. A detailed inspection of the maps reveals that Ti and V are strongly correlated showing identical distributions within the scan area. In most of the (Ti,V) grains Fe is present too. However, there are a few (Ti,V) grains that do not contain Fe and a few Fe grains that do not contain Ti or V. The maps indicate that the grain density is significantly higher in SPI-3 and that the grains itself are slightly smaller compared to AC ZYA. Assuming that grains can be detected up to a depth of 27~$\mu$m (see section~\ref{PIXEexp}) their density in AC ZYA can be estimated to about $6\times 10^6$~cm$^{-3}$.

It should be noted that using BE and EDX imaging in \cite{Sepioni47001} no grains were detected in SPI-3 at all, probably due to their small size and the inferior sensitivity of the methods used. Consequently, the sample was assumed to be free of Fe contamination \cite{Sepioni47001}, in clear contrast to our findings, see Tab.~\ref{Bulk_conc}.

\subsection{Analysis of single metallic grains}\label{PIXEgrains}
\begin{figure*}[ht!]
\centering
    \includegraphics[width=1\linewidth]{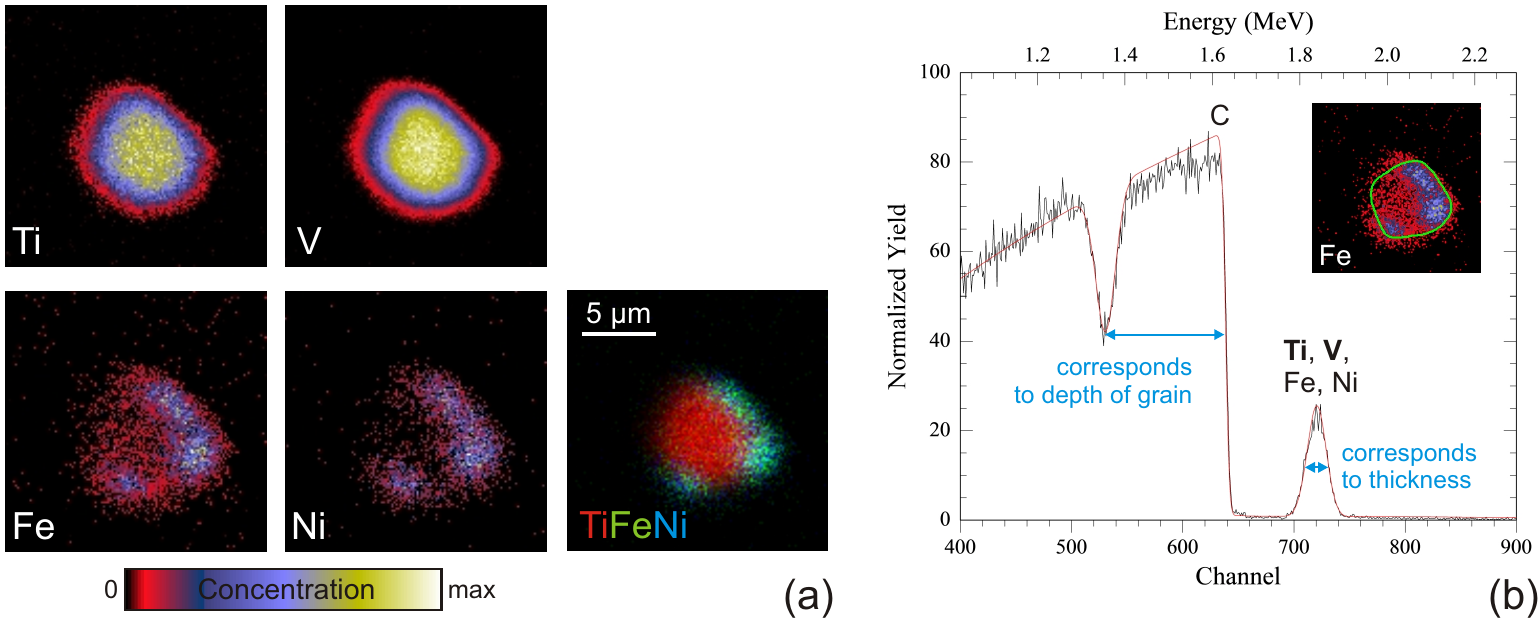}
    \caption{Ion beam analysis of a single grain in AC ZYA ($17.5~{\rm \mu m} \times 17.5~{\rm \mu m}$ scan area): (a) Element maps and composite map of Ti, Fe and Ni showing a homogeneous distribution of Ti and V  inside the grain and the location of Fe and Ni at the perimeter; (b) RBS spectrum extracted from the grain (the green curve in the Fe map represents the extracted scan area). From the XRUMP analysis (red curve) the metallic content, thickness and depth of the grain can be determined.}
    \label{zya_ac_c}
\end{figure*}

As can be seen in Fig.~\ref{PIXE_overview}, finding single grains of contamination is easy with ion beam microscopy. A careful analysis of these grains should allow to draw some conclusions on their magnetic properties. Therefore, single grains were selected from larger scan areas and each of the grains analyzed in detail using a smaller scan. Whereas PIXE allows to map elemental distributions with high sensitivity, RBS can be used to determine the depth below surface and thickness of the grain. The simultaneous recording of PIXE and RBS data, therefore, allows a reliable quantitative analysis in a non-destructive way, i.e. without the need of special sample preparation. As an example, Figs.~\ref{zya_ac_c}(a) and (b) show the PIXE and RBS analysis, respectively, of a single grain in the AC ZYA sample. As stated above, Ti and V have identical distributions, different from Fe and Ni that are located at the outside of the grain, both with a similar and irregular distribution along the grain's perimeter. This is also illustrated in the composite map of Fig.~\ref{zya_ac_c}(a) showing the distribution of Ti, Fe, Ni in red, green and blue color, respectively, with brighter colors referring to higher concentrations of the respective element.

Fig.~\ref{zya_ac_c}(b) shows the RBS spectrum extracted from the grain alone. The broad peak around channel 720 are protons backscattered from the metals where peak width and area reflect grain thickness and the total number of metal atoms, respectively. The grain is also visible as ``missing'' carbon in the dip around channel 530. From this dip position the depth of the grain can be determined. As quantitative analysis shows, this $7.0~{\rm \mu m} \times 5.5~{\rm \mu m}$ sized grain has a mass thickness of $(0.13 \pm 0.01)$~mg/cm$^2$, consists on average of 13.7\% Ti, 82.0\% V, 2.4\% Fe and 1.9\% Ni by weight and is located 4.35~$\mu$m below the graphite surface. It contains $(1.2\pm 0.1)$~pg Fe and $(0.96\pm 0.10)$~pg Ni. Assuming for simplicity that the grain is made of pure Vanadium with a mass density of $\rho=6.1$~g/cm$^3$, the geometrical thickness can be calculated to $d\approx 210$~nm. This and the analysis of other grains show that they are not spherical as assumed in \cite{Sepioni47001}, but flat disks that are oriented parallel to the graphene planes. This finding including the location of Fe at the perimeter of the grains agrees well with the EDX and TEM analysis of an AC ZYA sample reported in \cite{Venkatesan279}. Taking into account the high temperatures  $T>2000^\circ$C and pressures used in the production of HOPG from pyrocarbons \cite{Mikromasch}, this flat shape is to be expected. Metallic particles in the pyrocarbons, whatever their origin might be, melt at these temperatures and spread out perpendicular to the direction of pressure, i.e. perpendicular to the c-axis of the formed HOPG. Furthermore, the formation of iron carbides is to be expected under these conditions. Indeed, electron diffraction analysis showed that Fe in the grain is not metallic, but present as cementite Fe$_3$C \cite{Venkatesan279} in agreement with the findings in \cite{Tsuzuki2513,Vocadlo567,Talyzin57}.

\begin{figure}[ht]
\begin{center}
\includegraphics[width=1\columnwidth]{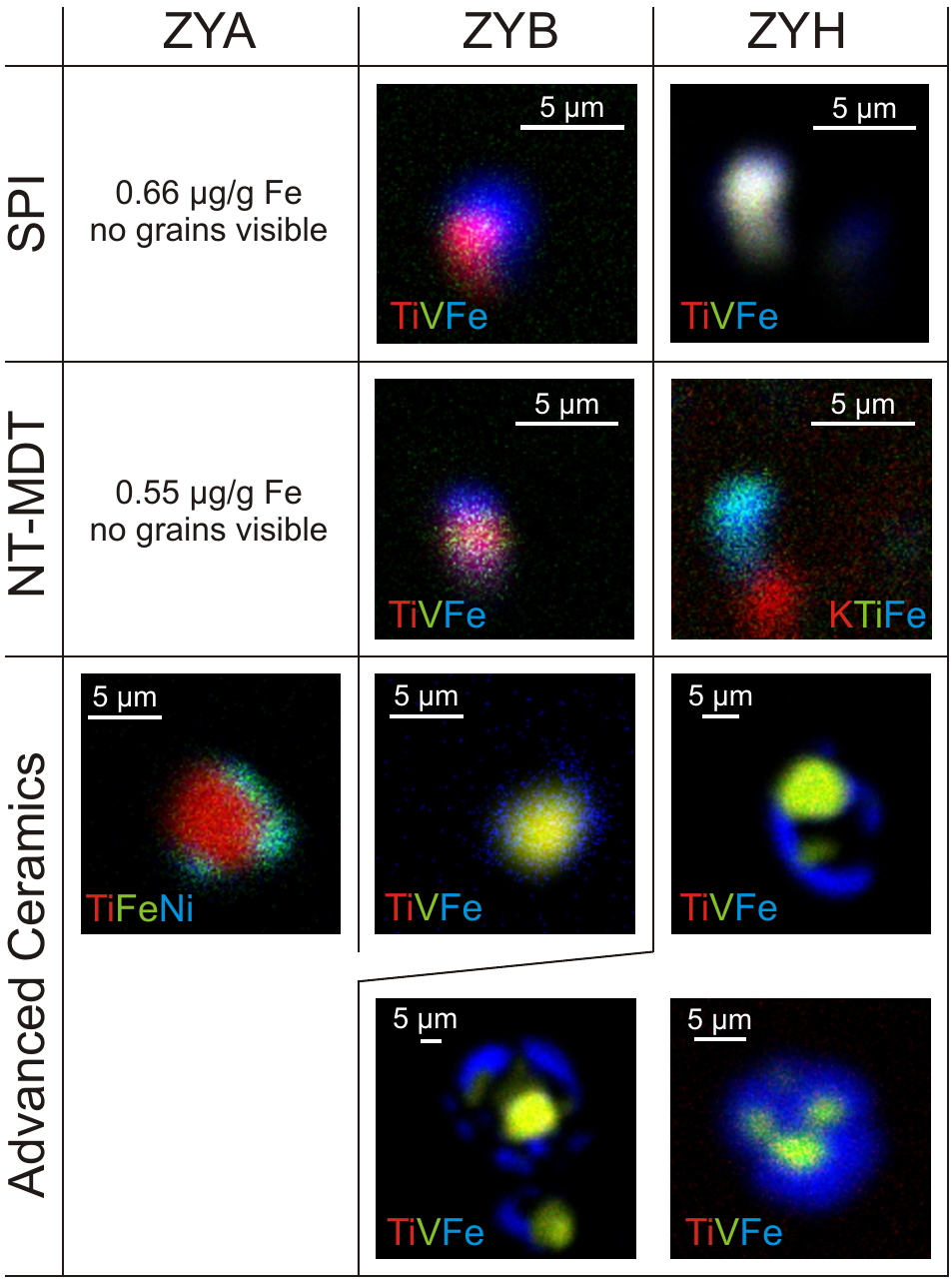}
\caption[]{Composite maps of single grains for all investigated samples except SPI-1 and NT-MDT ZYA where no grains could be found. The displayed elements and assigned colors are given in each map. As demonstrated, size and composition of the grains differ substantially between different samples, but also within the same sample (see AC ZYH).}
\label{PIXE_grains}
\end{center}
\end{figure}

Figure~\ref{PIXE_grains} shows composite maps of grains for all investigated samples with the exception of SPI-1 and NT-NDT ZYA where no grains could be detected within the detection limits of PIXE imaging. Indeed, these two samples are the only ones that do not contain Ti or V (see Tab.~\ref{Bulk_conc}). They do, however,
contain Fe of comparable amount as AC ZYA, presumably more or less homogeneously distributed within the bulk and not concentrated in grains as for the latter one. As can be seen from the maps, the grains differ substantially in size from 2~$\mu$m (NT-MDT ZYB) to $\approx 30$~$\mu$m (AC ZYH) and composition between different samples, but also within the same sample as is demonstrated for AC ZYH. Here, three grains of very different shape and size are displayed. In view of these variations, it is obvious, that the estimation of bulk concentrations from the analysis of a few single grains can lead to substantial errors. Indeed, taking the 1.2~pg Fe from the grain in Fig.~\ref{zya_ac_c} and the grain density of $6\times 10^6$~cm$^{-3}$ estimated from Fig.~\ref{PIXE_overview} one gets about 3~$\mu$g/g Fe as bulk concentration for AC ZYA where the true value is $(0.72\pm 0.07)$~$\mu$g/g. Obviously, the metal content of this grain is above the average compared to the other grains in AC ZYA.

\subsection{Comparison with EDX analysis}\label{c-edx}
\begin{figure}[ht]
\begin{center}
\includegraphics[width=1\columnwidth]{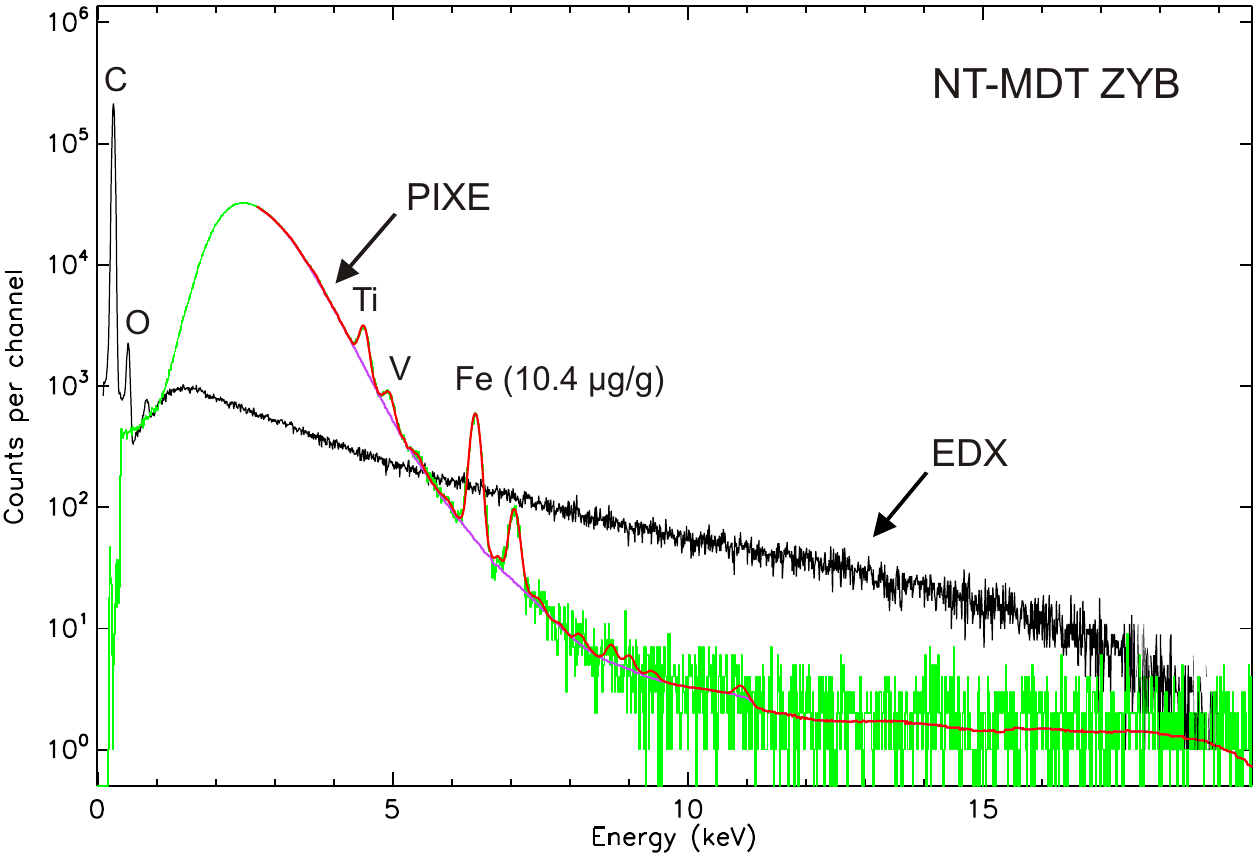}
\caption[]{EDX spectrum (black line) and PIXE spectrum (green line, together with GeoPIXE II fit, see Fig.~\ref{PIXEspec} for explanation) recorded on NT-MDT ZYB. In both cases, a large scan was made to obtain reliable bulk concentration values, however, no peaks for the trace elements can be detected in the EDX spectrum, demonstrating its insufficient sensitivity for trace element analysis in HOPG.}
\label{EDX_PIXE_comp}
\end{center}
\end{figure}

EDX is a wide-spread technique for elemental analysis and imaging and has been used recently in the characterization of contaminations in HOPG in \cite{Sepioni47001,Venkatesan279}. In order to compare its capabilities and limitations with those of ion beam microscopy some of the HOPG samples were studied using EDX. Figure~\ref{EDX_PIXE_comp} shows the X-ray spectra recorded from a NT-MDT ZYB sample using EDX and PIXE,
respectively, using a large scan area. Whereas the PIXE spectrum shows peaks for Ti, V and Fe, no peaks can be discerned for these trace elements in the EDX spectrum, despite Ti and Fe having concentrations $\gtrsim 10$~$\mu$g/g. As a detailed analysis shows, a typical MDL for Fe amounts to $\sim 200$~$\mu$g/g in EDX analysis, about a factor 1000 larger than for PIXE and far above the Fe bulk concentrations in all the HOPG samples. It is clear, that EDX cannot be used to measure the bulk concentrations of trace elements in HOPG directly.

\begin{figure}[ht]
\begin{center}
\includegraphics[width=1\columnwidth]{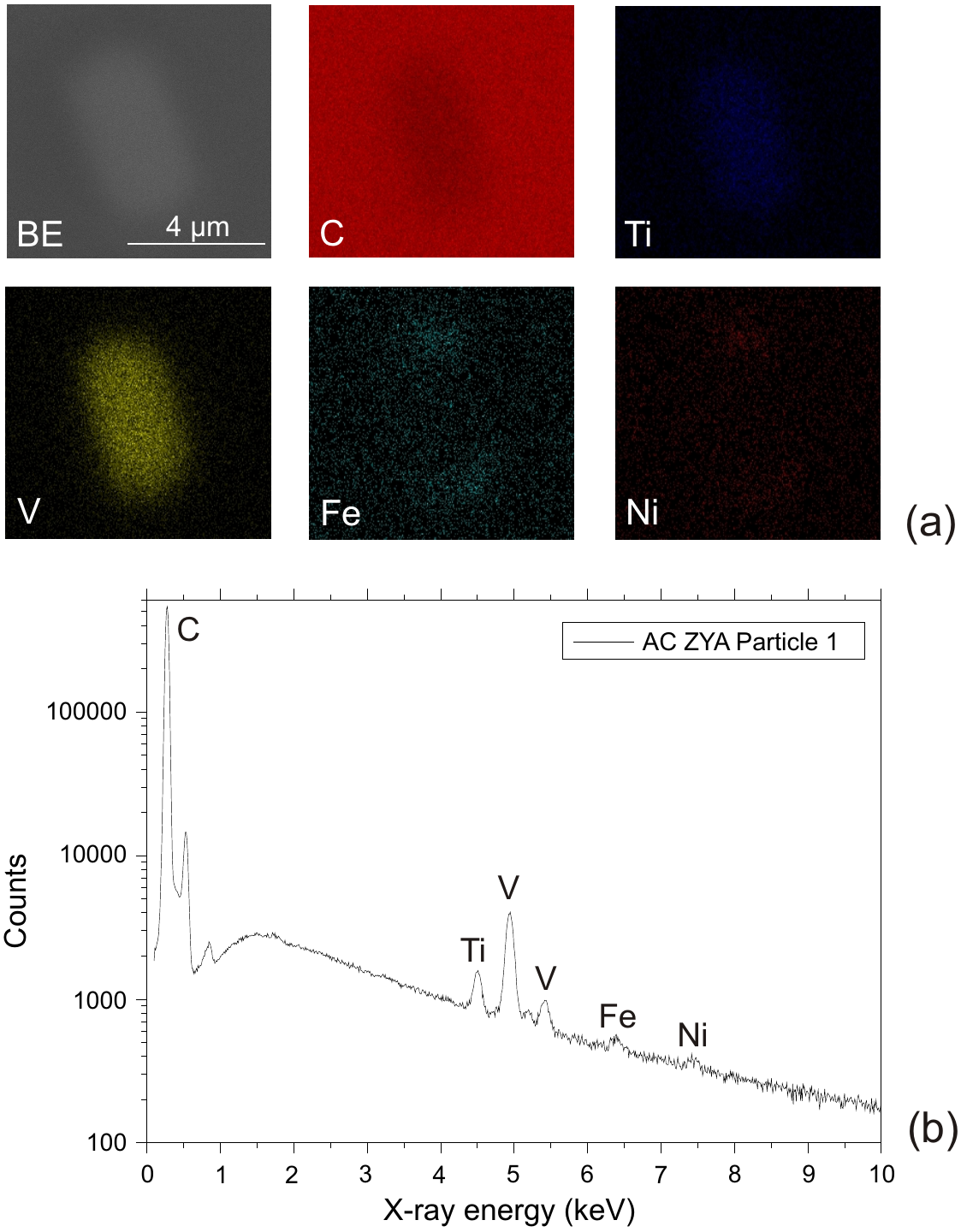}
\caption[]{EDX analysis of a single grain in AC ZYA: (a) BE and elemental maps of a $2.5~{\rm \mu m} \times 4.5~{\rm \mu m}$ sized grain. The Fe and Ni map indicate that both elements are enriched at the outer edge as already seen in Fig.~\ref{zya_ac_c}. (b) EDX spectrum showing peaks for the 3d-metals inside the grain. The peaks of Fe and Ni suffer from a comparably poor statistics.}
\label{EDX_AC_ZYA}
\end{center}
\end{figure}

In case of HOPG samples, where the trace elements are strongly concentrated in grains, EDX can at least be performed on single grains. Figure~\ref{EDX_AC_ZYA} shows such an EDX analysis on a grain in AC ZYA.
First, BE imaging is used to detect single grains directly below the graphite surface due to the $Z$-contrast in electron yield. Then, a small scan is made and the emitted X-rays are recorded (see Figs.~\ref{EDX_AC_ZYA}(a) for the BE and elemental maps and (b) for the EDX spectrum). From the spectrum, however, no direct information on depth and thickness of the grain can be obtained in contrast to RBS in ion beam microscopy, making a quantitative analysis, e.g. the determination of the metallic content, difficult. Qualitative analysis though shows that the Ti/V concentration ratio matches quite well the results from PIXE/RBS on the grain in Fig.~\ref{zya_ac_c} from the same sample, whereas the Fe and Ni concentrations are both twice as large as for PIXE. This might be due to differences in the composition of individual grains as pointed out earlier and/or due to the rather poor statistics in the EDX spectrum (see Fig.~\ref{EDX_AC_ZYA}(b)) and the insufficient knowledge of grain thickness and depth.

Assuming that the metallic content of single grains can be accurately determined with EDX, bulk concentrations can in principle be estimated from these data as was done in \cite{Sepioni47001,Venkatesan279}. This approach, however, has several weaknesses: (i) it requires that all the grains in a sample are comparable in metallic content and composition which is not necessarily the case as Fig.~\ref{PIXE_grains} shows; (ii) the number density of grains must be determined, e.g. using BE imaging. Since only a very low number of grains is present in HOPG in the near-surface area of $\lesssim 1$~$\mu$m depth even for large scan areas (in \cite{Venkatesan279} only three grains are visible for AC ZYA in a $1.0~{\rm mm} \times 0.9~{\rm mm}$ scan area!) substantial statistical errors exist; (iii) trace elements not enriched in grains are not taken into account at all. These difficulties lead us to the conclusion that EDX is useful for identifying and imaging metallic grains in HOPG, but cannot be considered as a reliable method for quantitative trace element analysis in graphite, in contrast to truly bulk-sensitive techniques as e.g. PIXE/RBS or INAA. Indeed, the Fe concentration in AC ZYA was estimated from SEM analysis (and magnetization data) to 6~$\mu$g/g in \cite{Venkatesan279}, whereas PIXE analysis always gave $< 1$~$\mu$g/g Fe for numerous of AC ZYA samples in the last ten years. This discrepancy might be due to the grain density which is stated to be about 0.25 per $100~{\rm \mu m} \times 100~{\rm \mu m} \times 0.5~{\rm \mu m}$ volume, i.e. $5\times 10^7$~cm$^{-3}$ for AC ZYA \cite{Venkatesan279}, about a factor eight higher than our estimations and what can be judged from the BE image in \cite{Venkatesan279} itself.

\section{Magnetic properties}\label{mp}
\subsection{Saturation magnetization and remanence}

As explained in the introduction, one of the aims of this work is to correlate the magnetization behavior of the HOPG samples with the one we can estimate taking into account the impurity concentration obtained by PIXE. In general, information on the impurity concentrations is mandatory in order to understand the origin of any unusual magnetic behavior of nominally non-magnetic samples. However, as we will point out below, the concentration
values alone are not sufficient to predict the behavior of the magnetization, just because the magnetic impurities in a graphite matrix can show different magnetic response upon several details, like their magnetic state (composition) and grain size. As an example, we note that Fe in graphite not always shows a ferromagnetic behavior \cite{pabloprb02}: a sample with an
inhomogeneous Fe concentration of up to 0.38\% (in weight) shows no magnetic order. If we implant single Fe atoms randomly distributed in a disordered graphite lattice one does not expect magnetic order, as confirmed after implanting Fe up to concentrations of $4000~\mu$g/g \cite{hoh08}.

\begin{figure}[ht]
\begin{center}
\includegraphics[width=1\columnwidth]{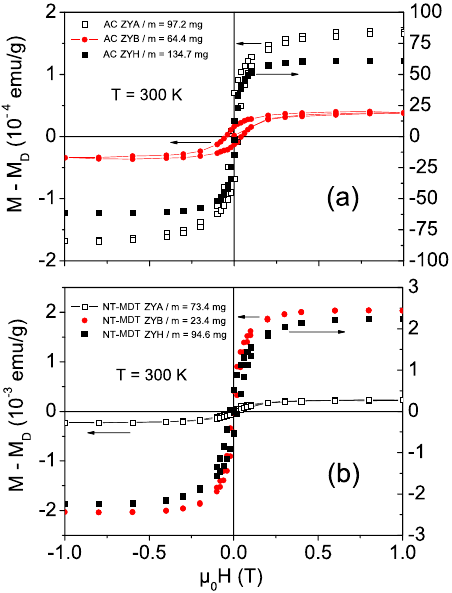}
\caption[]{(a) Magnetization at 300~K for the three AC samples. A linear diamagnetic background $M_{\rm D} = - 3.3 \times 10^{-3}~\mu_0H$emu/gT has been subtracted in all the three samples. (b) The same as in (a) but for the NT-MDT samples. The subtracted linear diamagnetic background was $M_{\rm D} = - 4.2 \times 10^{-3}~\mu_0H$emu/gT for the NT-MDT ZYA and NT-MDT ZYH samples and $M_{\rm D} = - 5.3 \times 10^{-3}~\mu_0 H$emu/gT for the NT-ZYB sample.} \label{Hhys}
\end{center}
\end{figure}

For the characterization of the magnetic behavior of the HOPG samples we measured the field and temperature hysteresis of the magnetic moment. In order to compare different samples with
different masses, the magnetization values given in this study are always normalized to the whole sample mass. Figure~\ref{Hhys}

shows the field hysteresis for six samples from two different companies at 300~K. The diamagnetic linear-in-field background has been already subtracted from the raw data (see caption for their values). Due to the parallel field direction, this background is relatively small and the ferromagnetic hysteresis is clearly recognized even without subtraction in most of the samples. From these hysteresis curves we obtain the magnetization values at saturation $M_{\rm sat} (\mu_0H = 1~$T) at different temperatures and the remanence, i.e. $M(\mu_0H = 0~$T) after sweeping the field back from $|\mu_0H| = 1~$T. For all samples we measured well
defined field hysteresis loops with weak temperature dependence.

\begin{figure}[]
\begin{center}
\includegraphics[width=0.9\columnwidth]{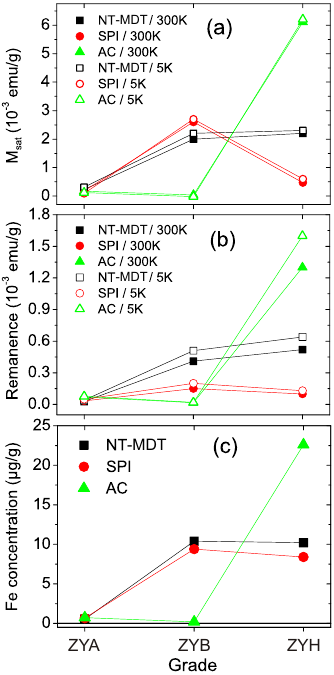}
\caption[]{(a) Magnetization at saturation at 300~K and 5~K for the nine samples vs. their grade. (b) Remanence vs. grade. The saturation and remanent magnetization have been obtained from the field hysteresis loops after subtraction of the diamagnetic background, see Fig.~\ref{Hhys} as example. (c) Fe concentration determined by PIXE vs. grade for all the samples. The lines are a guide to the eye only to allow a better comparison between the subfigures.}
\label{vsgrade}
\end{center}
\end{figure}

The magnetization values at saturation and at remanence at two temperatures as well as the Fe concentration are shown in Fig.~\ref{vsgrade} for the individual sample grades. This figure
reveals the following interesting details: (i) a very low Fe concentration below 1~ppm is found always for HOPG samples of ZYA grade (note that 10~$\mu$g/g Fe in graphite means about 2.1~ppm of Fe) with exception of AC samples where the lowest Fe concentration is found for the ZYB-grade sample and the sample with the highest Fe concentration ($\simeq 4.8~$ppm) is AC ZYH; (ii) at first glance, the magnetization data (Figs.~\ref{vsgrade}(a) and (b)) follow roughly the same trend as the Fe concentration. However, as we discuss below, a quantitative comparison of the magnetization values with the expected values from the measured Fe concentration added to the temperature dependence of the remanence indicate that the measured ferromagnetic response in some of the samples cannot be simply due to impurities.

\begin{figure}[]
\begin{center}
\includegraphics[width=0.9\columnwidth]{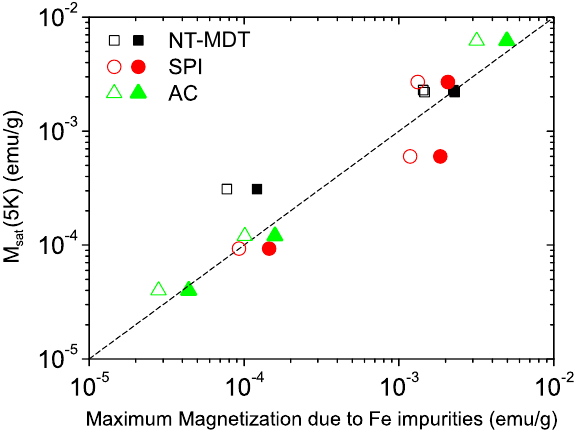}
\caption[]{Magnetization at saturation at 5~K vs. the maximum magnetization estimated assuming that all the Fe impurities would be ferromagnetic Fe (full symbols), ferrimagnetic magnetite Fe$_3$O$_4$ or cementite Fe$_3$C (open symbols), in double logarithmic scale. The dashed line indicates a one-to-one correspondence between the measured absolute values and the estimated ones.}
\label{corr}
\end{center}
\end{figure}

Figure~\ref{corr} shows the measured saturation magnetization at 5~K of all samples vs. the expected saturation value if \textit{all} Fe present in the sample would behave as pure
ferromagnetic Fe, Fe$_3$O$_4$ or Fe$_3$C, the latter one being the most likely case as discussed above. We recognize that a few samples show saturation magnetization values above the one estimated assuming a specific magnetic behavior for the Fe impurity. Let us compare the sample AC ZYH with sample AC ZYA. Assuming that $1~\mu$g/g of ferromagnetic Fe (alternatively as compounds Fe$_3$O$_4$ or Fe$_3$C) in graphite would produce a magnetization at saturation of $2.2 \times 10^{-4}~$emu/g ($1.4 \times 10^{-4}~$emu/g for both Fe$_3$O$_4$ and Fe$_3$C \cite{Stablein193,Talyzin57}), if \textit{all} the measured Fe would be ferromagnetic, for sample AC ZYH we would have the magnetization values at saturation of $5.0 \times 10^{-3}~$emu/g ($3.2 \times 10^{-3}$~emu/g) and for sample AC ZYA $1.6 \times 10^{-4}~$emu/g ($1.0 \times 10^{-4}$~emu/g), i.e. a ratio of $\simeq 32$ between the two samples. The ratio between saturation values at 5~K between those two samples is $\simeq 52$, i.e. about 62\% larger than the above estimated ratio. Nevertheless, and making the unrealistic assumption of the impurity in the samples being pure Fe, we would conclude that the measured Fe concentration roughly explains the absolute values as well as the difference in magnetization at saturation if \textit{all} Fe would be ferromagnetic.

This apparent correlation is less clear for other samples of different origin. For example, for the samples NT-MDT ZYH and NT-MDT ZYA the ratio between magnetization at saturation at 5~K is $2.3 \times 10^{-3} / 2.8 \times 10^{-4} = 8.2$, see Fig.~\ref{corr}, a ratio that does not agree with the expected ratio of 19 if \textit{all} Fe would contribute to the ferromagnetic signal. Furthermore, if we compare samples with similar Fe concentration but of different origins, i.e. the three samples with expected Fe magnetization around $1.5 \times 10^{-4}~$emu/g or the three around $2 \times 10^{-3}~$emu/g (see full symbols in Fig.~\ref{corr}), we recognize also that a large variation of the measured magnetization for similar Fe concentrations exists. A fact that should be not surprising since in general Fe is not homogeneously distributed in the micron-sized impurity grains as revealed by PIXE elemental imaging (see section~\ref{PIXEgrains}) and shown in \cite{Venkatesan279} for an AC ZYA sample. It is rather unlikely that it would provide the expected maximum magnetic moment at saturation for Fe, Fe$_3$O$_4$ or Fe$_3$C. In fact, the saturation magnetization for the NT-MDT ZYA sample, where no local Fe enrichment in the form of grains was found, is 2.3 or even 3.6 times larger than the highest expected saturation magnetization from Fe or Fe$_3$C, respectively. Even more, taking into account that Fe is most likely present in the form of Fe$_3$C, only the saturation magnetizations of the AC ZYA, SPI-1 and SPI-3 samples are still compatible with the observed impurity content, whereas the majority of samples shows magnetization values clearly above the possible contribution from all found magnetic impurities (note that apart from Fe, the maximum possible contribution of all other magnetic elements remains negligible, in comparison).

One might speculate that ferromagnetic nanoparticles, which can provide enlarged magnetic moments per impurity atom (e.g. Fe nanoparticles as small as 14~nm provide magnetic moments of 3~$\mu_{\rm B}$ per Fe atom \cite{barbara1991}), are responsible for the excess in magnetic moment compared to the impurity concentration. However, from our measurements we can rule out any significant contribution from such nanoparticles, e.g. features
like blocking temperature and superparamagnetism that would show up in the measurements if sufficiently small nanoparticles would be present in the samples were not observed. Furthermore, magnetism from Fe, magnetite or cementite nanoparticles is always characterized by a strong decrease of the saturation and remanent magnetization as well as coercive force with increasing temperature (see, e.g., \cite{ogrady1991}). However, the opposite
is observed in our samples. As was shown in  \cite{lipert2007}, even for large nanoparticles of up to more than 100~nm diameter and their agglomerates the coercive force at 300~K drops down to only 34\% of the value at 5~K, whereas for our samples the coercive force at 300~K still is 80\% of the 5~K value.

We note that in \cite{Sepioni47001} it was stated that no ferromagnetic signal and no impurity grains were found for SPI-2 and -3 in contrast to our findings. Furthermore, a large variation in saturation magnetization was reported between different NT-MDT ZYB and ZYH samples, again in contrast to our findings were the $M_{\rm sat}$ values as well as the Fe contents are nearly identical for both samples (see Fig.~\ref{vsgrade}). Interestingly, the NT-MDT ZYA sample showed the highest saturation magnetization among all samples in \cite{Sepioni47001}, despite having a Fe concentration about a factor 20 lower than the other samples -- according to our trace element analysis. One may tend to explain this clear deviation with strong variations in the impurity content between different batches of NT-MDT (and SPI samples as well). This, however, and from our experience with the HOPG samples from Advanced Ceramics, seems unlikely. It might, therefore, also indicate that no simple correlation exists between impurity content and magnetic properties -- as shown in this study.

In \cite{Venkatesan279} the saturation magnetization of the SPI-2 sample is about twice as large and the Fe content measured with INAA almost a factor three higher than for our sample. Since INAA is a suitable method for trace element analysis, the deviations between both findings might give an indication on the extent of variation between different batches of SPI-2 samples. However, since no information was given in \cite{Venkatesan279}  on sample cleaning and due to the lack of imaging capabilities in INAA analysis, it cannot be excluded that contaminations like the sideface contamination shown in Fig.~\ref{sideface}(a) have been overseen in \cite{Venkatesan279}. Nevertheless, assuming that Fe is present as Fe$_3$C in line with their own findings, the saturation magnetization is about 35\% larger than expected from the Fe concentration. More surprisingly, the saturation magnetization of the AC ZYA sample was found to be about a factor eight larger than in our study. From this value Venkatesan et al. estimated the Fe concentration in AC ZYA to be 6~$\mu$g/g \cite{Venkatesan279}, a value that clearly contradicts our measurements as discussed above. Whatever the contributions to this large magnetization are, it is far above the value that can be attributed to the impurity content inside the HOPG bulk itself.

From our findings presented above we conclude that in some of the HOPG samples, specially in those with low enough Fe concentration, an extra mechanism contributes to the observed magnetic order. This conclusion is similar to that obtained in \cite{pabloprb02}. As noted in the introduction, further support to this conclusion is obtained by the temperature dependence of the remanent magnetization discussed in the next section.

\subsection{Temperature dependence of magnetization}

Differences in the temperature dependence of the magnetization at saturation or at remanence can also provide a way to discern whether ferromagnetic contributions from magnetic impurities are at work in the samples. For all measurements done under a magnetic field, the temperature dependence of the total magnetic moment of a HOPG sample (in the case discussed here with field parallel to the graphene layers) is given by the sum of: (i) The intrinsic
Landau diamagnetism of the HOPG sample, which is given by a small misalignment between the field and the parallel to the graphene planes direction; (ii) The possible temperature dependence of the magnetic contribution from the substrate/sample holder, which in general should be negligible, if an appropriate holder is used; (iii) The temperature dependence of the ferromagnetic contribution itself. In the case of sample NT-MDT ZYB, for example, we measured $M(5\,$K$) = - 4.0 \times 10^{-3}~$emu/g and $M(300\,$K$) = - 3.2 \times 10^{-3}~$emu/g at a field of 1~T. We note that the overall change of the diamagnetic signal is $8 \times 10^{-4}~$emu/g, whereas $(2 \pm 1) \times 10^{-4}~$emu/g is the apparent change of the ferromagnetic contribution at saturation field. The uncertainty in the temperature dependence of the corresponding diamagnetic background is a non-negligible source of error when a
quantitative comparison of the $M_{\rm sat}(T)$ with appropriate models is required. We note that the difference in the temperature dependence between the magnetic moment at 2~kOe applied field and that at remanence of a HOPG sample, see Fig.~10 in Ref.~\cite{pabloprb02}, is not intrinsic but it is due to the constant diamagnetic background subtraction assumed for simplicity in that work. If one takes into account the weak temperature
dependence of the diamagnetic background at 2~kOe, an extra temperature dependent correction of $ \lesssim 5 \times 10^{-6}~$emu has to be taken into account,  enough to remove the difference between the two magnetic moment's temperature dependence.

\begin{figure}[ht]
\begin{center}
\includegraphics[width=1\columnwidth]{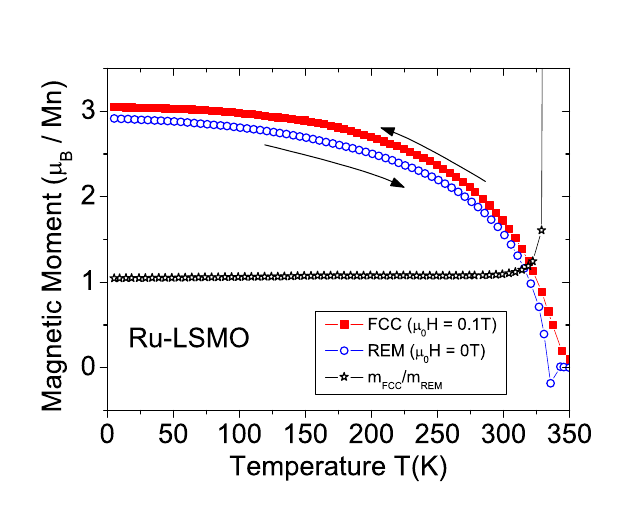}
\caption[]{Red squares show the magnetic moment (in units of Bohr magneton per Mn atom) of the 40~nm thick
Ru-doped LaSrMnO oxide film  measured by cooling in a field of 0.1~T (FCC).  The open blue circles
are the remanent magnetization measured at zero field by warming. The remanence is completely
irreversible, as expected. The black stars show the ratio between the magnetic moments; in this case
the same $y-$scale applies but unitless.}
\label{Ru-lsmo}
\end{center}
\end{figure}

Therefore, one choice is to compare the remanent magnetization measured at zero applied field. In case of the
ferromagnetism found in the HOPG samples the remanent (zero field) magnetization $M_{\rm rem}(T)$ can show a similar temperature dependence as the saturation one $M_{\rm sat}(T)$ (or at applied fields below saturation  as observed experimentally in \cite{barzola2}), at least in a temperature region clearly below the corresponding Curie temperature. This is expected when the applied field does not change the energy landscape of the domain walls and when there is no magnetic anisotropy that strongly changes with temperature. In general this is achieved applying the magnetic field parallel to an easy axis of the ferromagnetic material. The similarity between the temperature dependence of $M_{\rm sat}(T)$ and $M_{\rm rem}(T)$ can be observed in hard as well as soft ferromagnets. As an example of some ferromagnets where $M_{\rm rem}(T)/M_{\rm sat}(T) \lesssim~0.1$ and constant in the temperature region much below the Curie temperature, see \cite{san11}. As further example we show in Fig.~\ref{Ru-lsmo} the
magnetic moment of a Ru-doped LSMO ferromagnetic thin film at 0.1~T field applied parallel to the easy axis. In the same figure we show  the remanent magnetic moment measured, as usual, at zero field and on warming. These results indicate clearly that the ratio between both magnetic moments (stars) is constant within $\pm 1.5\%$ up to 300~K, below the Curie temperature $T_C \simeq 350~$K.

In the case of our HOPG samples let us take the measured values for the sample NT-MDT ZYB: At 5~K we have a
$M_{\rm rem} = 5.2 \times 10^{-4}$~emu/g and $M_{\rm sat} = 2.2 \times 10^{-3}$~emu/g; at 300~K these values are $4.1 \times 10^{-4}$~emu/g and $2.0 \times 10^{-3}$~emu/g, respectively. Note that the field was always applied parallel to the graphene planes of the samples, i.e. parallel to an easy axis. The ratios at 5~K and 300~K between the two magnetizations are 0.23 and 0.21 with an error $\gtrsim 0.01$. In other words, these results clearly indicate that the changes of both magnetization values with temperature are similar.

\begin{figure}[]
\begin{center}
\includegraphics[width=1\columnwidth]{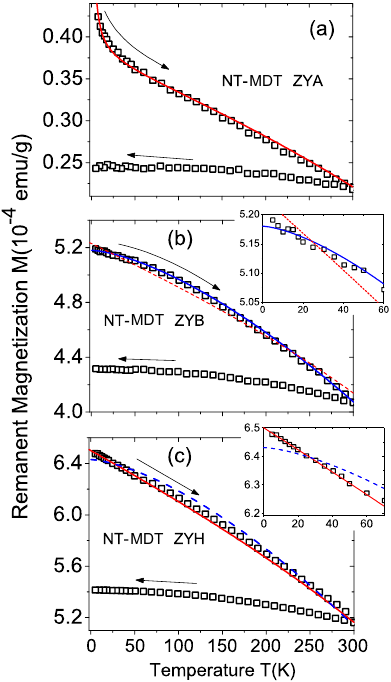}
\caption[]{Remanent magnetization (at zero field) vs. temperature for the three NT-MDT samples (after cooling them from 300~K to 5~K under a field of 1~T) on warming (5~K $\rightarrow$ 300~K) and cooling (300~K $\rightarrow$ 5~K). (a) Sample NT-MDT ZYA, the continuous red line was calculated from the sum of 2D-Heisenberg anisotropic spin wave model with the following parameters: critical temperature $T_c = 600~$K, spin-wave critical temperature due to low-energy spin-wave excitations $T_c^{SW} = 870~$K and anisotropy $\Delta = 0.001$ (see \cite{barzola2}) plus  $4.2 \times 10^{-5}/T$[emu K/g] due to a superparamagnetic contribution. (b) As in (a) but for the NT-MDT ZYB sample. The continuous blue line follows the case (i) described in the text with $n = 3/2$. For comparison the dashed red line follows the 2D-Heisenberg anisotropic spin waves model with parameters $T_c = 500~$K, $T_c^{SW} = 1700~$K, and $\Delta = 0.001$. The inset blows out the low temperature region. (c) Similarly for the NT-MDT ZYH sample: the continuous red line follows the 2D-Heisenberg anisotropic spin waves model with parameters $T_c = 550~$K, $T_c^{SW} = 1700~$K, and $\Delta = 0.001$. For comparison the dashed blue line follows the law for the case (i) with $n = 3/2$. The inset blows out the low temperature region.} \label{Thys}
\end{center}
\end{figure}

In Ref.\cite{barzola2} the temperature  dependence of the small ferromagnetic moment of graphite produced by proton irradiation has been determined with certain accuracy at different finite fields, subtracting the ``after irradiation'' signal minus the ``before irradiation'' signal, avoiding in this case arbitrary diamagnetic subtractions. Those results showed that the temperature dependence of the magnetization remains the same, independently of the applied field. Furthermore, we note that the changes in magnetization with temperature are relatively small, i.e. $10 \ldots 20\%$ within two orders of magnitude change in temperature. This indicates that the measured behavior occurs still far away from the Curie temperature and the number of domains does not change drastically with temperature. Second, the magnetically ordered regions are very probably relatively small in
size, with a relatively weak, if at all, magnetic anisotropy and/or pinning of domain walls. In this case, the change in $M_{\rm rem}(T)$ would be a thermally driven rotation of the magnetization vector, a process that is basically spin wave excitations. Obviously and due to its irreversible behavior, only the warming up curve of the  $M_{\rm rem}(T)$ curve can be compared with appropriate spin wave excitations models. As a function of temperature and by warming, the average magnetization vector goes through new potential minima that cannot be overwhelmed when the temperature is lowered, showing a temperature hysteresis typical for magnetically ordered materials, see
Fig.~\ref{Thys}.

We expect that different mechanisms will contribute to the remanence and influence its temperature dependence, depending whether large or small magnetic regions contribute. (i) For large enough ferromagnetic Fe particles or regions where due to the density of defects and their distribution within the layered structure the magnetization vector can be considered to be in a 3D potential, we expect to observe a $T$-dependence of the remanent magnetization compatible to excitation of spin waves  following the, e.g. 3D Bloch $T^{3/2}$ model. In this case we may have a law of the type $M(T) = M(0) ( 1 - CT^{3/2})$, where $C$ is a constant related to the spin wave stiffness and a fitting parameter. We note that this simple law applies only at $T \le 0.3~T_C$ ($T_C$ is the Curie temperature of the material) \cite{kitt}.

(ii) For samples with a weaker  3D ferromagnetic contribution, however, a quasi-linear temperature dependence for the remanent magnetization has been observed \cite{Esquinazi1156,dim13}. This dependence can be understood within the 2D Heisenberg anisotropic spin wave model. We note that this mechanism, first observed in proton irradiated HOPG samples \cite{barzola2} (see also \cite{xia08}) suggests that defects, whatever their origin, within
the graphene planes are responsible for triggering the observed magnetic order. The quasi-linear temperature dependence is an indication of excitation of 2D spin waves that reduce the magnetization linearly with $T$ \cite{dor61,lev92,ser93}. The 2D spin-waves magnetization follows as $ M(T)\simeq M^{sw}(T)M^I(T)$ obtained using perturbation theory techniques up to third order in spin waves \cite{bre76,lev92}; $M^{sw}(T)$ is the magnetization due to spin waves and $M^{I}(T)$ is due to an Ising model with the exchange renormalized by the spin waves, for more details see \cite{barzola2} and references therein.

(iii) Superparamagnetism is a possible third mechanism that can affect the temperature dependence of the remanent magnetization. According to \cite{mar96} it follows a simple $1/T$ dependence that can be added to the temperature dependence due to the other contributions. This contribution that has its origin in small enough ferromagnetic clusters, is expected to contribute mainly at low enough temperatures.

As an example of the different contributions we show  in Fig.~\ref{Thys} the results for the remanent magnetization of the NT-MDT samples measured after cooling the sample in a field of 1~T, where all three mechanisms can be observed. The observed
hysteresis between warming and cooling,  see Fig.~\ref{Thys}, is a clear evidence for the existence of a ferromagnetic state with Curie temperature above 300~K. The sample NT-MDT ZYA shows a behavior compatible with the sum of the contributions due to the mechanisms described in (ii) and (iii), i.e. a quasi-linear contribution plus a superparamagnetic state responsible for the low-temperature behavior (the value of the used parameters in the
fits are included in the figure caption), see Fig.~\ref{Thys}(a). Sample NT-MDT ZYB shows a temperature dependence that follows $\propto (1 - CT^{3/2})$  very probably in this case due to the large Fe impurities or disordered clusters. The observed behavior in this sample cannot be fitted with the 2D anisotropic Heisenberg spin wave model, independently of the chosen parameters. Figure~\ref{Thys}(b) and its inset show the calculated curves; one
can clearly realize, especially at low temperatures that the data do not follow the 2D anisotropic Heisenberg spin wave model. On the other hand, the behavior of the remanent magnetization of sample NT-MDT ZYH is compatible with the 2D anisotropic Heisenberg spin wave model, see Fig.~\ref{Thys}(c) and its inset, even though the amount of Fe contamination and the size of the grains are practically the same as in NT-MDT ZYB (see Tab.~\ref{Bulk_conc}
and Fig.~\ref{PIXE_grains} for comparison).

In spite of the above discussed quantitative and qualitative differences between the measured magnetic response in some of the HOPG samples, the apparent correlation between Fe impurity concentration and measured magnetization at saturation shown in Fig.~\ref{vsgrade} may leave a skeptical reader with the feeling that all the magnetic signal is due to impurities. We note, however, that in the way the HOPG samples are produced \cite{ina00}, there should be a correlation  between defect concentration (as well as hydrogen or other non-magnetic impurities) and the magnetic impurity concentration. The HOPG samples with less Fe impurity, grade ZYA, are obtained at larger annealing temperatures than those with grade ZYB or ZYH. Therefore, it should be not surprising that both, the Fe concentration and the defect density, are correlated. In fact, a direct correlation between the number of Fe atoms and defects, i.e.  an increase in the number of cation vacancies proportional to the Fe concentration, was already reported in TiO$_2$ \cite{nava2012,nava14}.

Independently done studies of the change of the magnetic signals after annealing at 2100$^\circ$C, a temperature well below the temperatures at which HOPG is produced, show a reduction of the saturation magnetization by a factor of two. This can only be explained by assuming that part of the magnetism observed is due to defects which can be removed by annealing and therefore supports the existence of a DIM contribution in HOPG samples of high grade and low impurity concentration \cite{mia12}. Finally, we note that
XMCD measurements at the near surface region of untreated HOPG samples revealed the existence of  magnetic order at the carbon K-edge, which is not related to any magnetic impurity\cite{ohldagnjp}. Those results support our main conclusion.

\section*{Conclusions}\label{conclu}
In this work we have done a complete trace element analysis using PIXE and RBS in different HOPG samples of different sources. The main  impurity that can contribute to the magnetic response of the samples is Fe, showing a maximum concentration of $\simeq 23~\mu$g/g for ZYH grade samples. The contamination at the sidefaces of as-received HOPG samples is notable and can exceed the one in the sample bulk. A thorough  cleaning is mandatory prior use of HOPG samples in contamination-critical applications. The analysis of single metallic grains indicates that they are not spherical but quasi flat disks oriented parallel to the graphene planes, in agreement with previous reports. The size and composition of the grains differ substantially between different HOPG samples but also within the same sample.

We have studied the elemental distribution of some of the samples with EDX in order to compare its capabilities and limitations with those of ion beam analysis. Our results clearly show that EDX cannot be used to measure  bulk concentrations of trace elements in HOPG with such a small impurity concentration directly but only and, to a certain extent, in single grains. Moreover, our comparative studies indicate that EDX cannot be considered a reliable method for quantitative trace element analysis in graphite, clarifying several weaknesses and discrepancies in the element concentration estimates done in the literature.

From the field and temperature hysteresis of the magnetic moment of the  HOPG samples we conclude that in some of the samples an extra contribution, other than those from magnetic impurities, to the observed ferromagnetic magnetization response exists. In agreement with previous reports this extra contribution is compatible with quasi two-dimensional defect-induced magnetism. The rough correspondence between the magnetization at saturation and the total Fe concentration indicates also that defect density and the impurity concentration can be correlated. No general answer, however, can be given even knowing the nominally magnetic impurity concentration, to the question whether magnetic impurities are or are not the reason for the observed magnetic response in a given HOPG sample.

\section*{Acknowledgement}
We gratefully acknowledge Francis Bern for providing us with his SQUID measurements
of Ru-doped LSMO film (Fig.11). This work was supported by the Deutsche Forschungsgemeinschaft
under contract DFG ES 86/16-1.

\bibliographystyle{model3-num-names}
% \bibliography{HOPG_Carbon_revised_version1}

\end{document}